\documentclass[12pt, nofootinbib]{article}
\usepackage{geometry} 

\usepackage{epsfig}
\usepackage{epstopdf}
\usepackage{amsmath}
\usepackage{bm}
\usepackage{caption}
\usepackage{tikz}
\tikzset{
  pics/carc/.style args={#1:#2:#3}{
    code={
      \draw[pic actions] (#1:#3) arc(#1:#2:#3);
    }
  }
}
\usepackage{sidecap}

\begin{document}

\title{Towards a realistic parsing of the Feynman path integral}
\author{K.B. Wharton\thanks{San Jos\'e State University, Department of Physics and Astronomy, San Jos\'e, CA 95192-0106} }

\date{} 

\maketitle
\abstract{The Feynman path integral does not allow a ``one real path'' interpretation, because amplitudes contribute to probabilities in a non-separable manner.  The opposite extreme, ``all paths happen'', is not a useful or informative account.  In this paper it is shown that an intermediate parsing of the path integral, into realistic non-interfering possibilities, is always available.  Each realistic possibility formally corresponds to numerous particle paths, but is arguably best interpreted as a spacetime-valued field.  Notably, one actual field history can always be said to occur, although it will generally not have an extremized action.  The most obvious concerns with this approach are addressed, indicating necessary follow-up research.  But without obvious showstoppers, it seems plausible that the path integral might be reinterpreted to explain quantum phenomena in terms of Lorentz covariant field histories.}

\section{Introduction}

When Feynman first developed the path integral formalism \cite{FPI} as a very different route to the predictions of quantum theory, it would have been plausible to wonder if this new approach might finally provide an answer to what is hapening between quantum measurements.  Specifically, it might have pointed the way towards a Lorentz-covariant description of an underlying reality based only upon spacetime-local entities.  (Such a description is what is meant by the word ``realistic'' in this paper; the lack of such an account has recently been termed the ``Lorentzian quantum reality problem'' \cite{Kent}.) 

The path integral comes with several promising features that give some hope to resolving the Lorentzian quantum reality problem.  Not only is the path integral based upon Lorentz-covariant components in spacetime (\textit{i.e.} the classical action $S$), but the mathematics bears a passing resemblance to that of statistical mechanics, wherein one sums over the probabilities of microstates to find the observable probability of their corresponding macrostate.  If applicable to quantum systems, this would have meshed with Einstein's view of the quantum state as analogous to statistical macrostates, and the path integral might have been viewed as a guidepost towards a hidden-microstate description of a spacetime-based reality.  

But today, after the path integral has been utilized and analyzed by several generations of physicists, such a hope seems remote.  True, most quantum foundations research has ignored the path integral in favor of the traditional Hamiltonian framework, but nevertheless there have been several major research programs aimed at better understanding the implications of the path integral.  These programs include Decoherent Histories \cite{Griffiths, GMH} and Quantum Measure Theory \cite{Sorkin, Dowker}.  Unless these programs have all made some systematic unnecessary assumption, it would seem that further analysis would be unlikely to uncover a solution to the Lorentzian quantum reality problem.

This paper will argue that such a systematic assumption has indeed occurred.  Granted, assumptions are necessary for \textit{any} quasiclassical analysis of the path integral, simply because most of the terms in the path integral are non-classical paths.  Reading any useful implication from this mathematics therefore requires some parsing of these infinitely-many terms into more classical-looking groups.  But the crucial point is that there are \textit{several} natural parsings that one might pursue: groups that look more like classical particle trajectories or groups that look more like classical field histories.   Following Feynman, every major research program in path-integral-interpretations has always chosen particles over fields, at least when interpreting single-particle path integrals.

Feynman's own leanings were clearly in the direction of particle trajectories.  In one presentation \cite{FeynVideo} he made it clear that he viewed the path-integral analysis as a particle theory, not a wave theory, using the particle viewpoint of the photon path integral to explain why electromagnetic fields are just a useful approximation.  Furthermore, much of his original motivation for the path integral was as an application to Direct Action electromagnetism \cite{DAEM} (in which the field is a fiction, and there are only particles).  Even though Feynman did give the wave perspective its due in general introductions to the topic, his deeper-level discussions of QED (both general and mathematical) were almost always couched in terms of particle paths.

Other modern research programs incorporate the very same bias.  In a recent extension of Decoherent Histories \cite{GMH}, with the explicit goal of finding a realistic underlying description, Gell-Mann and Hartle take the relevant parsing to be ``particle positions in the case of particles, four-dimensional field configurations -- both bosonic and fermionic -- in the case of quantum field theory, and histories of geometries and fields in the case of semiclassical quantum gravity.''  While this may sound like particles and fields are being given equal consideration, a closer look reveals that single-electron experiments are always parsed into particle trajectories; a field-based parsing is never even considered.

After a discussion of the relative merits of underlying classical fields vs. underlying classical particles in the next section, the following analysis will sketch out the basics of a field-based interpretation of the single-particle path integral.  Unlike the standard particle viewpoint, which can easily be shown to not yield a realistic interpretation, the first-order view of a field viewpoint reveals no obvious showstoppers for an eventual solution to the Lorentzian quantum reality problem.

\section{Fields vs. Particles}

Whether or not field viewpoints are \textit{superior} to particle viewpoints is not important; all that matters is that they are at least of comparable plausibility, in which case both perspectives should be considered.  To date, it is not clear that this has occurred.  And thanks to classical electromagnetism and general relativity, classical fields hardly need defense as a plausible realistic ontology.  

The case for fields in a quantum context is even stronger when one looks at path-integral formulations of quantum field theory, where every ``particle'' species actually corresponds to a quantum field, and in every case these fields have a strong classical analog.  For example, the classical Dirac-field analog for an electron \cite{Goldstein} allows for classical accounts of some of the curious behavior of spin-1/2 systems \cite{Ohanian, WSL}.

The particle assumption is particularly surprising in the case of single-photon experiments, as the closest classical analog to a photon is unarguably a classical electromagnetic field.  And most single-electron experiments have a nearly-perfect single-photon analog; almost any single-electron path integral problem can be reframed as a photon problem.  Applying particle-based logic to (say) a basic interferometer experiment would conclude that it is nonclassical for a photon to pass through both arms of an interferometer.  Indeed, this indication that there are ``two real paths'' is often the point at which particle-based interpretations of the path integral conclude that no realistic interpretation is conceivable.  And yet such behavior is \textit{entirely} classical in the context of electromagnetic fields, for which traveling down both arms of an interferometer is not only acceptable, but \textit{expected}.

While this example may bring to mind realistic fields, there are certainly plenty of other examples that bring to mind realistic particle trajectories.  For example, one can send a photon through a single beamsplitter followed by detectors on each output port.  This leads to an end result (one detector firing) that is certainly far more reminiscent of a classical particle than it is a classical field.  

But notably, all such examples only seem like particle behavior at the very beginning and the very end of the experiment -- at preparation or measurement.  And the beginning and end of experiments have a special role when using path integrals; they are precisely the points at which boundary conditions are imposed by the path integral mathematics.  Since these boundary conditions have to be imposed \textit{regardless} of how one parses the integral, that constraint can make \textit{anything} look like a particle at that special point, even fields.  (Consider a field constrained to be emitted from a local point; it would look like a dispersed field everywhere except at the constraint.)  The initial and final constraints on the path integral, then, might imply that it is unwise to restrict ones attention to classical particle trajectories based on phenomena evident at these special points alone.

And even in this beamsplitter example, there is a classical field history that can explain the outcome.  Every beamsplitter not only has two output ports, but also two input ports.  If a known wave is heading towards a beamsplitter, another unknown electromagnetic wave could be incident on the other \textit{input} port.  The two fields could then classically interfere in a way that would force all the field onto single output port.  While this would be a dubious dynamical explanation, it is certainly a valid classical field history, violating no physical law except the empirically-motivated Sommerfeld radiation condition \cite{Price0}.  (And even the empirical justification for this condition cannot be extended down to single-photon intensities, due to the uncertainty principle.) 

These examples have shown that there is every reason to consider classical field histories, even when interpreting single-particle path integrals.  The following sections will now demonstrate the promise of such a viewpoint.

\section{The Single Particle Path Integral}

For any experiment where a single particle undergoes two consecutive position measurements at spacetime locations $(\bm{x}_0,t_0)$ and $(\bm{x}_1,t_1)$ respectively, the path-integral formulation of quantum theory predicts that the unnormalized joint probability distribution over all pairs of possible positions (keeping the measurement times fixed) is

\begin{equation}
\label{eq:PI}
P(\bm{x}_0,\bm{x}_1)=\left| \sum_{\bm{x}_0\to\bm{x}_1} exp(i S/\hbar) \right|^2.
\end{equation}

Here the ``sum" is the infinitesimal limit of a discretized set $\bm(Q)$ of all possible spacetime trajectories from $(\bm{x}_0,t_0)$ to $(\bm{x}_1,t_1)$, and $S$ is the classical action of the particle on that trajectory.  Typically, one is more interested in conditioning these probabilities on the known value $\bm{x}_0=\bm{x}_i$ and then computing the conditional probability density over the actual outcome $\bm{x}_1=\bm{x}_f$.  This is easily accomplished via the usual conversion between joint and conditional probabilities:

\begin{equation}
\label{eq:norm}
P(\bm{x}_f|\bm{x}_i)=\frac {P(\bm{x}_i,\bm{x}_f)}{\int_{-\infty}^{\infty} P(\bm{x}_i,\bm{x}_1) d\bm{x}_1}.
\end{equation}

This procedure leads to an automatically-normalized probability distribution over $\bm{x}_f$, and is  provably equivalent \cite{FPI} to the probabilities predicted by Hamiltonian quantum mechanics.  Furthermore, this framework is more evidently time-symmetric than standard dynamical QM, because here the time-symmetry is automatically built into the formalism rather than just a special consequence.  (Especially when couched in terms of joint probabilities, all path-integral predictions must necessarily be as symmetric as the underlying Lagrangian density that produces the action $S$; no such inherent time-symmetry can be generically deduced from Hamiltonian-evolution combined with Born-like rules that generate conditional probabilities.)   

There are further advantages and generalizations of the path integral, but the remainder of this paper will focus on the above two-position-measurement framework.  The central question is whether or not the above mathematics might allow a realistic interpretation, in which exactly one intermediate spacetime-based history can always be said to occur.

\subsection{The Statistical Framework}

The path integral does not posit any law-like dynamics (or even stochastic dynamics) that takes initial states to later states; all paths seem to initially be on an equal footing.  This brings to mind the analysis of instantaneous states in statistical mechanics.

In statistical mechanics (absent dynamics), one considers a space of possible microstates $\mu$, consistent with known information, and assigns a (usually equal) probability $P(\mu)$ to each allowable microstate.  This allows one to deduce the probability of any observable macrostate by simply calculating $\sum P(\mu_i)/Z$, where $\mu_i$ is the subspace of microstates consistent with that macrostate, and $Z$ is the partition function $\sum P(\mu)$.  As this mathematics bears a passing resemblance to equations (\ref{eq:PI}) and (\ref{eq:norm}), statistical mechanics might be viewed as a framework for a realistic interpretation of the path integral.  

The attraction of this classical analysis should be clear.  In the case of statistical mechanics, there is always one real microstate, and the fact that it is not known naturally leads to a probabilistic description.  Absent this precise knowledge, multiparticle descriptions of mechanical systems encode known correlations using a high-dimensionality configuration space, but there is no school of thought that takes such a space to be real (ontological).  In statistical mechanics, this state is clearly a state of knowledge (epistemic), underpinned by one real microstate that exists in ordinary three-dimensional space.

\subsection{One Real Trajectory?}  

A naive extension of this statistical approach to the path integral might note that the sum in (\ref{eq:PI}) is over trajectories, and extend this logic to a ``one real trajectory'' interpretation.  These trajectories would then lie in four-dimensional spacetime, and one would expect to calculate probabilities by summing over all such trajectories consistent with the endpoints (which would map to the observable macrostate).  

But even though this endpoint-constrained sum does indeed appear in (\ref{eq:PI}), there are three obvious mathematical barriers to such a realistic interpretation.  The first is that the term being summed over, $exp(i S/\hbar)$, can be negative, and therefore cannot be associated with a probability of a trajectory.  This breaks the analogy to $P(\mu)$.  The second problem is closely related, in that this term is not even a real number.  The third and final problem is that the probabilities generated by (\ref{eq:PI}) do not result from a simple sum of probabilities, but a square of the total.  Every trajectory therefore contributes to the observable probabilities, and one cannot simply dismiss all-but-one of them as not being ``real".

It is perhaps not widely appreciated that the latter two problems can be easily solved in an elegant manner, simply by changing the kinematical possibility space.  This is closely related to the central point of this paper; the choice of what one considers a ``realistic'' history can dramatically change ones assessment of this mathematics.  This partial-solution to the above problems (originally due to Sinha and Sorkin \cite{Sinha}) is explained in detail in the next subsection.

\subsection{Two Real Trajectories?}

By explicitly squaring the sum, it is a trivial matter to rewrite equation (\ref{eq:PI}) in terms of pairs of trajectories, path $A$ and path $B$, that both go from the same $\bm{x}_0$ to the same $\bm{x}_1$.

\begin{eqnarray}
\label{eq:PI2}
P(\bm{x}_0,\bm{x}_1)&=&\sum_{A\in\bm{Q}}  exp(i S_A/\hbar) \sum_{B\in\bm{Q}} exp(-i S_B/\hbar) \notag \\
&=& \sum_{(A,B)\in\bm{R}} exp[i(S_A-S_B)/\hbar].
\end{eqnarray}
Here $S_A$ is shorthand for the action of the particle on trajectory A, etc.  Again $\bm{Q}$ is the set of all paths from $\bm{x}_0\to\bm{x}_1$, and now $\bm{R}$ is the set of all path-pairs.  Notice how the square of (\ref{eq:PI}) has been subsumed into the expanded kinematics of (\ref{eq:PI2}).  

This expression would be unchanged if $A$ and $B$ were swapped, because there is no pair of trajectories for which the reverse pair does not appear in the original sum.  Doubling this expression by adding the $A\leftrightarrow B$ version (and then dividing by two) then yields a purely real expression:
\begin{equation}
\label{eq:PI3}
P(\bm{x}_0,\bm{x}_1) = \sum_{(A,B)\in\bm{R}} \cos\frac{S_A-S_B}{\hbar} .
\end{equation}
The total sum here is non-negative, because any given path appears in both $A$ and $B$.  For terms where any given path appears twice, one gets $\cos(0)=1$, and these positive values will at least cancel any potentially-negative cross-terms.  (Indeed, this was all derived from (\ref{eq:PI}), which is explicitly positive.)  

Unfortunately, this does not lead to a ``two real trajectories" interpretation of the path integral, for the sole remaining reason that the individual terms in (\ref{eq:PI3}) can be negative.  Therefore, they cannot be probabilities.  Perhaps this is not surprising, as this is also the essential problem with assigning realistic probabilities to Wigner functions in phase-space formulations of quantum mechanics.  Nevertheless, having resolved some of the earlier problems in this manner, the choice of kinematics might be seen as crucial to whether or not a realistic interpretation can be found.  

It should be noted that \textit{if} the terms in (\ref{eq:PI3}) were somehow never negative, a realistic intermediate account would be available -- but it would imply a very curious description of how a particle gets from $\bm{x}_0$ to $\bm{x}_1$.  After every measurement (at point $\bm{x}_0$), this mathematics would indicate that the particle would split into two pieces, and these pieces would take usually-different paths to the same destination $\bm{x}_1$.  Such a conclusion would raise significant foundational questions about these new ``half particles''.

This is roughly how far Quantum Measure Theory \cite{Sorkin} has analyzed realistic interpretations of the path integral before introducing fundamental changes to logic and probability (in an effort to interpret (\ref{eq:PI3}), especially in light of three-path interference experiments).  This paper proposes that such a dramatic step should not be taken unless all other avenues are exhausted.  

For example, if three-path interference is so problematic, why not postulate an ontology consisting of three real paths?  From there, one would quickly get forced to 4-path, 5-path, many-path ontologies, which seems to be unexplored territory.  Perhaps these avenues have not been explored because many-path ontologies already appear so strange, as indicated by the ``half particle'' discussion.  However, later we will see how many-path ontologies can be reinterpreted as a single field, a far more familiar construct.

\subsection{Many Real Trajectory-Pairs}

In the previous subsection we saw the importance of choosing the kinematical possibility space, and how different choices can solve (some) seemingly intractable interpretation problems.  With this in mind, one can consider splitting the sum in (\ref{eq:PI3}) into distinct pieces, $P(\bm{x}_0,\bm{x}_1) =\sum_i(F_i+G_i)$, where each term $F_i$ or $G_i$ itself contains a sum over \textit{many} trajectory pairs.  This, at least, has the correct form of the standard statistical framework.

Furthermore, negative probabilities can be eliminated if the groupings are chosen such that $F_i>0$ and $G_i=0$.  In other words, $F_i$ is found by taking some subset $a_i$ of possible trajectory pairs $(A,B)\in a_i\subset \bm{R}$, and summing $cos[(S_A-S_B)/\hbar]$ over these path-pairs.  By design, the result must be strictly positive ($F_i>0$).   Similarly, $G_j$ is derived from another subset $b_j\subset\bm{R}$ of possible trajectory pairs, where the corresponding sum over $(A,B)\in b_j$ is exactly zero.  As long as every pair of particle trajectories is represented in one of these sets, these expressions will be mathematically identical to the ones above (and the $G$ terms will all cancel).
\begin{equation}
\label{eq:sum}
P(\bm{x}_0,\bm{x}_1) = \sum_{i} \left( \sum_{(A,B)\in a_i} \cos\frac{S_A-S_B}{\hbar} \right).
\end{equation}

It is a logical certainty that such a parsing must always be possible, as (\ref{eq:PI}) indicates that $P(\bm{x}_0,\bm{x}_1)$ is positive.  The $G_j$ terms disappear by design; they sum to zero.  A similar parsing is (implicitly) assumed by some path-integral analyses in order to get rid of the wildly non-classical paths.  Those paths are far away from the $\delta S=0$ extremization condition for classical particles, and with $\hbar$ non-zero, such paths have phases that almost always exactly cancel.  These ``non-possibilities'' can then be lumped into a group like $G_j$ and ignored.  The only new step here is to note that the negative amplitude terms can always be grouped together into strictly positive partial sums.

The obvious conclusion of this analysis is that the kinematical possibility space could be assigned to the collection of path-pairs $a_i$ as a whole.  If this step is made, then a realistic interpretation of the path integral is indeed possible.  In every experiment, one particular set of path-pairs $a_i$ really happens between measurements, but it is never clear which one, even after the final measurement.  Due to this uncertainty, one assigns a probability 
\begin{equation}
\label{eq:pfi}
P(a_i) = F_i = \left( \sum_{(A,B)\in a_i} \cos\frac{S_A-S_B}{\hbar}  \right) >0.
\end{equation}
 to each distinct set, and then sums over all possibilities $\sum_i F_i$ and normalizes just as in statistical mechanics.  This gives the correct joint probability for the ``macrohistory'' $(\bm{x}_0,\bm{x}_1)$, even though there is only one real ``microhistory'' $a_i$.
 
This proves that it is indeed possible to have a realistic (statistical mechanics-like) interpretation of the path integral, but at the expense of having the fundamental possibility space consist of \textit{many} particle trajectory-pairs, not merely one.  Interpreting such a realistic account would seem to require a quite unusual perspective on what is happening between measurements (much stranger than even the half-particle trajectories discussed at the end of Section 3.3), but there is at least an existence proof that such an account is possible.

However, there is still room for improvement, because there are still so many ways in which any net positive sum like (\ref{eq:PI3}) can be parsed into sets of path-pairs $a_i$ and $b_i$.  The next subsection will demonstrate that a further-restricted parsing can reframe the relevant kinematical space as disjoint sets of single paths.

\subsection{Many Real Trajectories}

Although the previous section technically supplies a way to realistically parse the path integral, one concern is how to make sense of a many-path-pair ontology.  The most problematic case is when a given single-path appears in multiple different sets $a_i$, in different combinations with other paths.  To eliminate this concern, consider a different parsing of (\ref{eq:PI3}), into sets of \textit{single} paths, $c_i\subset\bm{Q}$.  Every path in the original sum (\ref{eq:PI}) must show up in one set $c_i$, and only one set.  Then (\ref{eq:PI3}) can be rewritten as
\begin{equation}
\label{eq:PI4}
P(\bm{x}_0,\bm{x}_1) \!=\!\!\! \sum_i \sum_{A\in c_i} \left( \sum_{B\in c_i} \cos\frac{S_{\!A}-S_{\!B}}{\hbar} + \!\! \sum_{B\notin c_i} \cos\frac{S_{\!A}-S_{\!B}}{\hbar}\right).
\end{equation}

It turns out to be relatively simple to force the second term in this expression -- the one for which the path pair $(A,B)$ spans two different sets $c_i$ -- to always sum to zero.   (For now, one can leave aside the outer sum over the different sets $i$, just considering the sums over $A\!\in\!c_i$ and $B\!\notin \!c_i$.)  Setting $\hbar=1$ for simplicity, this second term will necessarily sum to zero so long as
\begin{equation}
\label{eq:term}
\sum_{A\in c_i} \cos S_A \sum_{B\notin c_i} \cos S_B + \sum_{A\in c_i} \sin S_A \sum_{B\notin c_i} \sin S_B =0.
\end{equation}

Because every path in $\bm{Q}$ is exactly one set $c_i$, this can be rewritten in terms of only paths in the set $c_i$ and two overall constants $C\equiv\sum_{B\in\bm{Q}} \cos S_B$ and $D\equiv\sum_{B\in\bm{Q}} \sin S_B$:
\begin{equation}
\label{eq:Cond}
\sum_{A\in c_i} \cos S_{\!A} ( C\,-\!\sum_{A\in c_i} \cos S_{\!A} ) + \!\!\sum_{A\in c_i} \sin S_{\!A} ( D\,-\!\sum_{A\in c_i} \sin S_{\!A} )=0.
\end{equation}
Enforcing this condition for each set $c_i$ will therefore eliminate cross-set interference.  For example, any group of paths for which both $\sum_{A\in c_i} \cos S_A=0$ and $\sum_{A\in c_i} \sin S_A=0$ would be permitted.  This is not an available option for \textit{every} set $c_i$, because then $C$ and $D$ would both have to be zero, but there seems to be no reason why a general parsing could not use many sets that fall into this simple category.  (These sets would then correspond to the unimportant terms $G_i$ from Section 3.4.)  

The constraint (\ref{eq:Cond}) on each given set $c_i$ could even be relaxed further, by reintroducing the sum over different sets ($i$) from (\ref{eq:PI4}).  This would replace (\ref{eq:Cond}) with a single interrelated constraint between all sets; either way, the second term in (\ref{eq:PI4}) could be ignored when calculating probabilities, and it would simplify down to
\begin{equation}
\label{eq:PI5}
P(\bm{x}_0,\bm{x}_1) \!=\!\!\! \sum_i \left( \sum_{A\in c_i} \sum_{B\in c_i} \cos\frac{S_A-S_B}{\hbar}\right).
\end{equation}
This double-sum is a familiar expression from section 3.3, so we can undo the ``squaring'' analysis of that section to recover the simple and promising
\begin{equation}
\label{eq:PI6}
P(\bm{x}_0,\bm{x}_1) \!=\!\!\! \sum_i  \left| \sum_{A\in c_i} exp(i S_A/\hbar) \right|^2.
\end{equation}

This almost looks like the original path integral formula, but thanks to the above parsing of possibilities into groups of paths $c_i$, it now is represented as a linear sum over strictly positive terms!   Clearly, a classical probability interpretation is again available: between measurements, one set of paths $c_i$ is always taken (subject to the constraint (\ref{eq:Cond})), but it is unclear which one.  The probability of each set is now calculated via a familiar method, using \textit{only} the paths that appear in the set $c_i$. 

Note that the constraint  (\ref{eq:Cond}) does not eliminate interference; it only eliminates interference between distinct-possibility sets.  Interference is still evidently possible between paths \textit{within} a given set.  And because of the appearance of many-path interference in well-established phenomena (say, triple-slit experiments), there is no escaping the conclusion that each set $c_i$ must contain many paths.  There is no way to reduce $c_i$ down to a single path or two, in agreement with the earlier analysis.

The remaining task, then, is to somehow find a natural way to interpret each many-path set $c_i$ as a distinct physical possibility.  The particle story is clearly unwieldy; this would imply that each particle splits up into many pieces at $\bm{x}_0$, only to recombine at $\bm{x}_1$.  Alternatively, one could consider that these path-sets $c_i$ are a representation of a single, spacetime-valued field.

 \section{Paths to Field}
 
The preceding analysis proves that it is possible to assign a classical-ignorance interpretation to the path integral, essentially no different than the framework of classical statistical mechanics as described in Section 3.1.  The most straightforward reading of Eqn. (\ref{eq:PI6}) is that there are many non-interfering possibilities that may occur, and each single possibility corresponds to some set of trajectories $c_i$.  The probability of each trajectory-set can be assigned a positive value, and one of these sets can be said to ``really happen''.  Just as in statistical mechanics, the other possibilities are only needed to normalize the (subjective) probabilities one assigns to $c_i$, not for any interference-like process.  (Interference may happening with a given set of trajectories $c_i$, but there is no longer any intereference \textit{between} sets, because the positive probabilities of each set combine as a simple classical sum.)

The only key difference between this reading and classical statistical mechanics is that here one is assigning probabilities to entire microhistories rather than instantaneous microstates.  Still, given that it might require the combination of \textit{many} particle trajectories to parse the path integral into the attractive form (\ref{eq:PI6}), what physical microhistory might the set $c_i$ represent?   For example, a high order reflection off of a diffraction grating seems to have many separated paths that are interfering.  In such cases, a given set $c_i$ incorporating this interference is necessarily associated with a group of trajectories spanning a sizable region of space.  The most straightforward interpretation of such a ``many path" set $c _i$ would then be that of an extended field, not that of a subdivided particle. 

If this assessment is correct, the task is then to map the relevant parameters of the set $c_i$ onto a continuous, spacetime-valued field.  One template for how this might work (albeit in reverse) is the connection between the standard complex wavefunction and the collection of possible paths that particle might take in Bohmian mechanics.\cite{dBB}  Here the density of paths maps to the field amplitude, and the direction of the paths (at any given point) maps to the gradient of the phase.  Although this map requires that the paths never cross, it may be the case that one can always build each set $c_i$ out of non-intersecting paths, allowing a similar map to work here.  

Note that even if this Bohmian-style approach was successful, it would not simply reproduce the Bohmian interpretation of the standard wavefunction.  The most obvious difference, for a given set $c_i$ corresponding to some final boundary constraint $\bm{x}_1$, is that all of the paths come together at the point where the particle is measured.  The corresponding complex field, then, would continuously build up in amplitude as it approaches this measurement point, unlike the standard wavefunction.  A closer connection might be drawn to a recent two-boundary approach to Bohmian mechanics \cite{Sutherland}, but even here there are several key differences at the ontological level. 

If it turned out that intersecting paths were sometimes required in order to preserve the condition (\ref{eq:Cond}), there are certainly other ways to map a set of paths onto a spacetime-valued field.  One simple extension would be to average the path-velocities at a given point to recover the phase gradient; another would be to allow some small finite number of overlapping paths, with each set corresponding to different field values.  (Instead of a two-component complex field, one could utilize a 4- or 8-component field as well.)  The crucial point is that there is no need to recover the standard quantum wavefunction $\psi$, as the probabilities are already built into the above framework.  What one is aiming to recover here is a realistic field-based description of what is happening between measurements, and this will certainly not look like $\psi$ (at least not at measurement).

For a rough outline of what this field-interpretation might amount to, consider the example of a free particle sequentially detected at $\bm{x}_0$ and $\bm{x}_1$ -- perhaps a photon exchanged between atoms at these locations. One set of possible paths that connect these locations is given by the thin solid lines in Figure 1; we can call this set $c_1$.  In this case, the lines are straight except for at the plane equidistant from $\bm{x}_0$ and $\bm{x}_1$.  Interpreting these lines as representing the phase gradient of a field, the lower half would correspond to a spherical wave diverging from $\bm{x}_0$, and the upper half would correspond to a spherical wave converging onto $\bm{x}_1$.  The discontinuity in the slope of the lines at the equidistant plane would have to correspond to a (spatially non-uniform) phase shift in this field, converting the diverging wave into a converging wave.

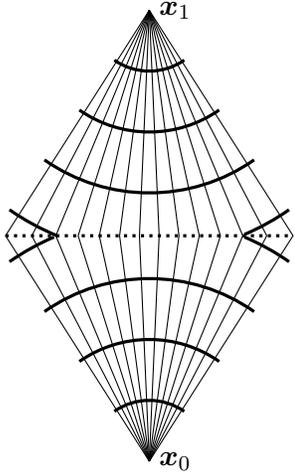
\begin{SCfigure} [2.5]
\label{fig:Fig1}
\centering
\begin{tikzpicture}[scale=3]

\draw (0,0) --(0.043660906,1) --(0,2);
\draw (0,0) --(0.131652385,1) --(0,2);
\draw (0,0) --(0.221694469,1) --(0,2);
\draw (0,0) --(0.315298505,1) --(0,2);
\draw (0,0) --(0.414213174,1) --(0,2);
\draw (0,0) --(0.520566535,1) --(0,2);
\draw (0,0) --(0.637069587,1) --(0,2);
\draw (0,0) --(-0.043660906,1) --(0,2);
\draw (0,0) --(-0.131652385,1) --(0,2);
\draw (0,0) --(-0.221694469,1) --(0,2);
\draw (0,0) --(-0.315298505,1) --(0,2);
\draw (0,0) --(-0.414213174,1) --(0,2);
\draw (0,0) --(-0.520566535,1) --(0,2);
\draw (0,0) --(-0.637069587,1) --(0,2);

\draw [very thick] (0,0) ++(55:.27) arc (55:125:.27);
\draw [very thick] (0,0) ++(55:.54) arc (55:125:.54);
\draw [very thick] (0,0) ++(55:.81) arc (55:125:.81);
\draw [very thick] (0,0) ++(55:1.08) arc (55:67:1.08);
\draw [very thick] (0,0) ++(113:1.08) arc (113:125:1.08);

\draw [very thick] (0,2) ++(-55:.27) arc (-55:-125:.27);
\draw [very thick] (0,2) ++(-55:.54) arc (-55:-125:.54);
\draw [very thick] (0,2) ++(-55:.81) arc (-55:-125:.81);
\draw [very thick] (0,2) ++(-55:1.08) arc (-55:-67:1.08);
\draw [very thick] (0,2) ++(-112:1.08) arc (-112:-125:1.08);

\draw [very thick, dotted] (-0.637069587,1) --(0.637069587,1) ;
\node [right] at (0,0) {$\bm{x}_0$};

\node [right] at (0,2) {$\bm{x}_1$};

\end{tikzpicture}

\caption{A representation of one possible way that atoms at $\bm{x}_0$ and $\bm{x}_1$ might exchange a photon.  The thin lines comprise a single set of path integral trajectories, corresponding to one particular set $c_1$.  These trajectories are straight except for at the mid-plane (dotted line), and can be jointly reinterpreted as a single field (such that the trajectories are normal to the phase fronts).  The phase fronts of the corresponding field are noted with thick lines.  In this case, the phase of the field must be discontinuous at the mid-plane, where the original trajectories change slope.}

\end{SCfigure}

Given that it has an intermediate phase anomaly, the field in Figure 1 does not solve any deterministic wave equation, so it cannot result from an action-extremized Lagrangian.  This is a general feature of such intermediate fields: there is typically no solution to Maxwell's equations subject to two localized constraints.  But the whole point of the path integral mathematics is that non-action-extremized paths are still relevant, so it would be quite strange to dismiss this field solution on the grounds that it was not action-extremized in its own right.  (An equivalent argument would be to dismiss the thin lines in Figure 1 as being ``nonclassical'' because they bend in the middle; it hardly seems surprising that nonclassical trajectories would correspond to a nonclassical field.)  The key point, however, is that even with an intermediate phase anomaly, the field in Figure 1 can still provide a localized (spacetime-based) account of what is happening between the emission and absorption of the electromagnetic energy.

\begin{figure} 
\label{fig:Fig2}
\centering
\begin{minipage}{.47\textwidth}
\centering

\begin{tikzpicture}[scale=3]

\draw (0,0) --(0.02202276	,0.504404552) --(0.022120189,1.493363943) --(0,2);
\draw (0,0) --(0.067606299,0.51352126) --(0.06853295,1.479440115) --(0,2);
\draw (0,0) --(0.115990108,0.523198022) --(0.118744751,1.464376575) --(0,2);
\draw (0,0) --(0.168259656,0.533651931) --(0.1741191,1.44776427) --(0,2);
\draw (0,0) --(0.225813579,0.545162716) --(0.236494304,1.429051709) --(0,2);
\draw (0,0) --(0.290531459,0.558106292) --(0.308454612,1.407463616) --(0,2);
\draw (0,0) --(0.365046842,0.573009368) --(0.393797768,1.38186067) --(0,2);

\draw (0,0) --(-0.021641476,0.495671705) --(-0.021548209,1.506464463) --(0,2);
\draw (0,0) --(-0.064137424,0.487172515) --(-0.063325122,1.518997536) --(0,2);
\draw (0,0) --(-0.106141058,0.478771788) --(-0.10393471,1.531180413) --(0,2);
\draw (0,0) --(-0.148297647,0.470340471) --(-0.144025907,1.543207772) --(0,2);
\draw (0,0) --(-0.191261944,0.461747611) --(-0.184215269,1.555264581) --(0,2);
\draw (0,0) --(-0.235739635,0.452852073) --(-0.225125437,1.567537631) --(0,2);
\draw (0,0) --(-0.2825358,0.44349284) --(-0.267424407,1.580227322) --(0,2);
\draw (0,0) --(-0.332617792,0.433476442) --(-0.311871022,1.593561306) --(0,2);
\draw (0,0) --(-0.38720373,0.422559254) --(-0.359373573,1.607812072) --(0,2);
\draw (0,0) --(-0.447895243,0.410420951) --(-0.411071821,1.623321546) --(0,2);
\draw (0,0) --(-0.51688764,0.396622472) --(-0.46845946	,1.640537838) --(0,2);

\draw [very thick, dotted] (-0.51688764,0.396622472) --(0.365046842,0.573009368) ;
\draw [very thick, dotted] (-0.46845946,1.640537838) --(0.393797768,1.38186067) ;

\draw [very thick] (0,0) ++(55:.27) arc (55:145:.27);
\draw [very thick] (0,0) ++(55:.54) arc (55:75:.54);
\draw [very thick] (0,0) ++(125:.54) arc (125:145:.54);

\draw [very thick] (0,2) ++(-55:.27) arc (-55:-145:.27);
\draw [very thick] (0,2) ++(-55:.54) arc (-55:-80:.54);
\draw [very thick] (0,2) ++(-132:.54) arc (-132:-145:.54);

\draw [very thick] (-0.51,0.549) --(0.37,0.637);
\draw [very thick] (-0.5,0.87) --(0.375,0.87);
\draw [very thick] (-0.485,1.1885) --(0.38,1.102);
\draw [very thick] (-0.47,1.504) --(0.385,1.333);
\node [right] at (0,0) {$\bm{x}_0$};

\node [right] at (0,2) {$\bm{x}_1$};

\end{tikzpicture}

\captionof{figure}{Another possible set of paths corresponding to the same observed events as Figure 1, only this time the corresponding field has two distinct phase anomalies.}
\end{minipage}%
\hfill%
\begin{minipage}{.47\textwidth}
\centering
\label{fig:Fig3}

\begin{tikzpicture}[scale=3]
\begin{scope}
\clip(-1.5,-0.1) rectangle (3,2.1);

\draw [very thick] (0,0) ++(45:.25) arc (45:135:.25);
\draw [very thick] (0,-0.3) ++(65:.8) arc (65:115:.8);
\draw [very thick] (0,-0.7) ++(72:1.45) arc (72:108:1.45);

\draw [very thick] (-.45,1) --(.45,1);

\draw [very thick] (0,2.7) ++(-72:1.45) arc (-72:-108:1.45);
\draw [very thick] (0,2.3) ++(-65:.8) arc (-65:-115:.8);
\draw [very thick] (0,2) ++(-45:.25) arc (-45:-135:.25);

\draw[thin] (0,0) to [out=87.5,in=-87.5] (0,2);
\draw[thin] (0,0) to [out=82.5,in=-82.5] (0,2);
\draw[thin] (0,0) to [out=77.5,in=-77.5] (0,2);
\draw[thin] (0,0) to [out=72.5,in=-72.5] (0,2);
\draw[thin] (0,0) to [out=67.5,in=-67.5] (0,2);
\draw[thin] (0,0) to [out=62.5,in=-62.5] (0,2);
\draw[thin] (0,0) to [out=57.5,in=-57.5] (0,2);
\draw[thin] (0,0) to [out=52.5,in=-52.5] (0,2);
\draw[thin] (0,0) to [out=47.5,in=-47.5] (0,2);
\draw[thin] (0,0) to [out=42.5,in=-42.5] (0,2);

\draw[thin] (0,0) to [out=92.5,in=-92.5] (0,2);
\draw[thin] (0,0) to [out=97.5,in=-97.5] (0,2);
\draw[thin] (0,0) to [out=102.5,in=-102.5] (0,2);
\draw[thin] (0,0) to [out=107.5,in=-107.5] (0,2);
\draw[thin] (0,0) to [out=112.5,in=-112.5] (0,2);
\draw[thin] (0,0) to [out=117.5,in=-117.5] (0,2);
\draw[thin] (0,0) to [out=122.5,in=-122.5] (0,2);
\draw[thin] (0,0) to [out=127.5,in=-127.5] (0,2);
\draw[thin] (0,0) to [out=132.5,in=-132.5] (0,2);
\draw[thin] (0,0) to [out=137.5,in=-137.5] (0,2);

\node [right] at (0,0) {$\bm{x}_0$};

\node [right] at (0,2) {$\bm{x}_1$};

\end{scope}

\end{tikzpicture}

\captionof{figure}{Another possible set of paths corresponding the same observed events as Figure 1, only here the corresponding field has a continuously-distributed set of phase anomalies.}
\end{minipage}
\end{figure}

Figure 1 merely represents one set of paths; other distinct sets are possible as well.  Figure 2 demonstrates a different set (say, $c_2$) where the corresponding field undergoes two phase anomalies.  Figure 3 shows yet another set ($c_3$), where the paths are curved; this indicates a continuous-phase-anomaly in the corresponding field.  Crucially, only one of these sets need be ascribed any ``reality''; given the condition (\ref{eq:Cond}), there is no net interference between these sets, just within each set.  The probability of each corresponding field is positive, and the probabilities sum linearly, as shown in (\ref{eq:PI6}).  

For this last example (Figure 3) one can also imagine filling the space between the atoms with some continuously-varying index material, such that each of the thin-line paths in this particular set $c_3$ corresponds to a term with exactly the same action $S_A$.  In this case, one would certainly expect the corresponding probability of $c_3$ to be maximum  in (\ref{eq:PI6}), dominating all other path sets.  And notably, the corresponding field in Figure 3 would then be the classical field expected in this exact situation: the wave would literally be focused onto the point $\bm{x}_1$ by the intermediate material.  (One can also convert Figure 1 and Figure 2 into such ``classical'' scenarios by putting appropriate lenses on the dotted lines.)

In general, if there \textit{is} a large set of paths with the same action (or actions that differ by integral multiples of $2\pi\hbar$, as for a diffraction grating), the overwhelming probability of such a set points to an overwhelming probable intermediate field configuration -- the very ``classical'' field that would traditionally be used to describe such cases.  If this is not the case, and there is no ``classical'' field,  then there is no one obvious intermediate field solution.  Nevertheless, the above analysis shows that in these cases there are always \textit{many} possible intermediate field solutions, it just happens that none of them look particularly classical (due to phase anomalies).  Summing over the probabilities of each of these intermediate options yields the precise joint probability that the field goes from $\bm{x}_0$ to $\bm{x}_1$, as per (\ref{eq:PI6}).

These examples are mostly meant to be illustrative; a deeper quantitative analysis is certainly required to make further progress.  But for now, the remainder of this paper will focus on larger, big-picture questions, to see whether such further effort might plausibly be thought to succeed.  

\section{Discussion}

The previous sections raised various unresolved technical questions, including how to generally parse the paths into sets that obey condition (\ref{eq:Cond}), and exactly how to map each set of paths to a field.  But the largest potential concerns at this point are likely not these technical details, but instead big-picture issues that would have to be addressed to turn this proof-of-principle into a full-fledged interpretation.  Some of these concerns have evident solutions; others will require substantial work in order to resolve.  These are now addressed in turn:

\textbf{What about multiple particles?} The path integral approach naturally scales to multiple particles, where the action $S$ is now simply a function of all of the particles, not just one.  Instead of summing over only one set of paths $\bm{Q}$, one jointly sums over $\bm{Q}_1$, $\bm{Q}_2$, etc. for each particle.  (There is also a natural, classical way to introduce configuration-space constraints on a collection of such paths.)  In this multiparticle case, $c_i$ would refer to a set of paths for \textit{each} particle, where the paths for each particle would correspond to a different field.  (Entanglement issues will be addressed shortly; for now note that multiparticle path-integral analysis naturally gives the correct probabilities even for Bell-inequality violating scenarios, as explicitly shown in \cite{Sinha, WMP}.)  

\textbf{What about identical particles?}  Having a different field for each particle is awkward for identical particles.  For example, it would seem that the collection of all photons should map to something comparable to the classical electromagnetic field, not a different field for each photon. (Certainly, this would be natural in the high-field limit.)  Substantial work remains to be done to show whether or not this goal is attainable.  Fortunately, the path integral does not sum over different path-permutations for identical particles, so there is not an essential difference between the set of paths for a single particle and the (larger) set of paths for many identical particles.  It seems plausible that if one starts with an ontology consisting of many paths that map to a single field, increasing the number of particles on those paths need only correspond to increasing the energy density of the very same field configuration, because interference between distinct sets $c_i$ has already been forced to zero. 

\textbf{Why is the probability of a field configuration related to a sum of paths?}  One might expect that for an ontology where only the field is real (not particles on paths), one should be able to find the probability rule directly from the proposed field configuration instead of the paths $c_i$.  Since we know that anomaly-free field histories map to the high-field classical limit (say, Maxwell's equations), the probability of a given field history should be inversely related to some measure of the size of its nonclassical phase anomaly.  Zero-anomaly cases would be the most likely (recovering classical field theory), but since that is generally not an option, \textit{some} anomaly is usually forced by the boundary constraints.  It seems reasonable that fields with smaller anomalies (via some measure) should be more probable than fields with large anomalies.

Looking at Eqn. (\ref{eq:PI5}) with this point in mind, it seems evident that this comparison of every path-pair in $c_i$ is one way to (inversely) quantify the phase anomaly of the corresponding field.  Consider that the unitless action $S/\hbar$ acts like a phase in quantum theory, and the $S$ that appears in this equation is the \textit{classical} action (without anomalies).  Comparing the phase between all pairs of paths, as in $\sum\cos[(S_A-S_B)/\hbar]$, would therefore supply an inverse-measure of the minimum phase anomaly needed to make the total field coherent at the final boundary constraint.  (This term is obviously maximized if no anomaly is needed, if every representative path is in phase.)  Ideally, one might even find some master principle to more carefully justify this supposition, in analogy to the fundamental postulate of statistical mechanics.

Indeed, simple anomaly-based models of quantum phenomena have already been developed.\cite{WhartonInfo,LOQT,Schulman}  In these approaches, some intermediate anomaly takes the initial condition to the final condition, and a measure of this anomaly determines the probability of the entire history.  If a full field-based version of these models was successfully developed, and was compatible with the above arguments, this would have crucial implications for both quantum foundations and quantum field theory.  

\textbf{There is no single consistent way to group the paths.}  Even though any given future measurement constraint has a set of paths that can be parsed into realistic-possibility sets $c_i$, different future measurements will lead to different sets.  For example, in a double-slit experiment, a position measurement immediately after the slits (before interference could occur) would only have sets $c_i$ in which every path passed through the same slit.  But delaying the time of the position measurement (after potential interference) would lead to sets $c_i$ where the paths passed through both slits before converging onto the measurement point.  If each $c_i$ is interpreted as a field, the former case would yield a realistic field history that really did pass through only one slit, while the latter case would yield a realistic field history that really did pass through both slits.

This issue has been discussed in explicit detail in \cite{WhartonInfo}, where it is shown that this dependence on the future measurement is not a problem but instead a crucial feature of the ``Lagrangian Schema'' framework already assumed by the path integral.  True, if one tried to translate this account back into the standard ``Newtonian Schema'', where the future is determined by initial conditions and (possibly stochastic) dynamical laws, it looks quite strange.  But the Feynman path integral is predicated on a different sort of logic, where entire histories are examined ``all at once'', and those histories are constrained in both the past and future.  Taking away the ability for the future measurement to constrain the past is equivalent to denying the applicability of the path integral in the first place, because the future boundary constraint is a mathematical necessity for any path-integral-style analysis.

\textbf{Doesn't any spacetime-based account of entanglement imply nonlocal causation, voiding the goal of Lorentz invariance?} This class of concerns arises from various quantum no-go theorems, as well as examination of how entanglement works in cases such as Bohmian mechanics \cite{dBB}.  But here again, the role of the future constraint (built into the path integral mathematics) resolves the problem.  Given a future constraint that can at least be partially chosen by an external experimenter (\textit{i.e.} the type of measurement to perform, or when to perform it), any past events that are constrained by that future setting explicitly violate the independence assumptions behind every quantum no-go theorem.\cite{WhartonInfo}  (Formally, this falls under what is usually termed the ``retrocausal loophole''.)  

While it is still true that there is a net-nonlocal influence in such future-boundary models, it is not the direct, instantaneous influence that one sees in Bohmian mechanics.  Instead, it is closer to a ``zigzag'' influence, always on contiguous time-like paths through spacetime (see \cite{PriceWharton} for a clear recent discussion.)  These influences can indeed be Lorentz invariant, as nowhere do they require a direct spacelike interaction.  And even this ``zigzag'' picture should probably be set aside in favor of a  Lagrangian-style, all-at-once analysis.  As carefully described in \cite{WhartonInfo}, entanglement scenarios with future boundaries can be framed in terms of a knowledge-updating agent rather than any literal transmission at the ontological level.  If learning about a local constraint tells one something about the global possibility space (i.e. informs an agent which field configurations are possible elsewhere), then updating the relevant probabilities can have the same apparent effect as quantum steering.  If all possible field configurations have a Lorentz-covariant description, and only one actual configuration happens, this problem is resolved.

\textbf{Measurement is still treated as a special interaction.}  From the above discussion, it is clear that the future measurement must literally be imposed as a boundary constraint in order to make this analysis work.  But this might raise concerns similar to the usual ``measurement problem'' in ordinary quantum theory.  Specifically, the concern here is that some interactions are treated differently from others, with some ill-defined ``measurements'' imposed as boundary conditions while other interactions are not.  

In standard QM, this problem is indeed serious, because the decision of whether to collapse the wavefunction or to entangle it with the measurement apparatus leads to two incompatible descriptions, with no obvious middle ground.  The most intractable aspect of this problem comes about because entangled states live in a large configuration space that typically can't be mapped onto the separable results of a collapse.  

But for this one-real-field parsing of the path integral, configuration space plays the same epistemic role as in classical statistical mechanics.  Any spacetime-based description of a large region that includes a system and a measurement apparatus can \textit{always} be split up into equivalent separate descriptions of the system, the device, and the correlations between them.

Having taken this deep conflict between spacetime and configuration space off the table, there seem to be plenty of ways to resolve the remaining issues.  In classical physics there are many examples of a ``middle ground'' between boundary constraints and interactions -- say, electromagnetic fields interacting with conductors, or lower-entropy systems interacting with higher-entropy thermal reservoirs.  In both of these cases, one ends up with essential differences depending on whether the interacting systems have a comparable number of possible states.  This offers a potential resolution for the path integral: interactions between vastly different-entropy systems can naturally be imposed as effective boundary constraints on the smaller entropy system, while interactions between comparable systems can merely imply correlations.  For further discussion and some technical examples, see \cite{LOQT}.

\section{Conclusions}

The most important result of the above analysis is a proof-of-principle demonstration that the Feynman path integral is formally compatible with a realistic, Lorentz-covariant account of what is happening between measurements.  Crucially, this can be done without having to modify probability theory in any way, unlike most other path-integral-motivated research \cite{GMH,Sorkin}.  

The fact that this simple result has remained unnoticed for so long is perhaps attributable to the fact that most foundational research has been focused on Hamiltonian-based versions of quantum theory, and the exceptions have all implicitly assumed that evident single-particle behavior must be explained in terms of realistic particles.  If one uses realistic fields to explain such behavior, a standard probabilistic account can seemingly be recovered: one Lorentz-covariant field history occurs between any two measurements, and adding the (positive) probabilities of each possible field yields the correct joint probability.

Extensive work remains to turn this proof-of-principle into a full-fledged interpretation, especially where it comes to measurements other than position.  But before such work begins, it is certainly worth taking a step back and considering what this result might mean for the ``sum over all paths'' concept in the first place.  True, this well-known prescription (combined with the usual rules for summing and squaring amplitudes) does indeed seem to give the correct quantum results for a single particle.  But if we can better reframe the ontology in terms of a realistic field, perhaps we have been drawing the wrong lessons from the single-particle path integral.  Indeed, it is the application of ``sum-over-everything'' to field theory that causes so many mathematical difficulties: the lack of a natural measure on the space of field histories, the infinities and renormalization problems that ensue in that domain, etc.  

It is therefore notable that the above analysis implies that the relevant possibility space might be much smaller.  Many of the above sets of paths $c_i$ can correspond to zero-probability cases, and need never be summed over at all.  (The question of \textit{which} paths end up in such zero-probability sets depends on the particular future measurement, but as discussed in the previous section, this is a benefit of the ``Lagrangian Schema'' implied by the path integral mathematics.)  Furthermore, each probability would be associated with a possible (spacetime-based) field history, with no mathematical implication that many contradictory events were somehow happening at once.  As in classical statistics, one could not normalize these probabilities without considering the full space of all allowable histories, but so long as the probabilities of the wildly-anomalous histories dropped off fast enough, there would be no need to include them.  This raises the intriguing possibility that there's no need to sum over \textit{all} field configurations -- the very step that leads to so many technical problems.  

The biggest implication for quantum foundations would be the availability of an explicit between-measurement account, entirely in terms of spacetime-local fields.  This would also be true for entanglement experiments: no ``real'' configuration space would be required, justifying the $\psi$-epistemic view of entangled states. \cite{WhartonInfo,PriceWharton}  Such a framework might even provide a new and generally unexplored path to quantum gravity.  After all, block-universe field-history accounts are evidently more compatible with general relativity (GR) than are canonical quantization approaches where time plays a special role.  Instead of trying to force GR's spacetime into QM's configuration space, one could instead hope to use this approach to bring quantum phenomena back into ordinary spacetime.  

In conclusion, by simply changing our perspective on the kinematical space of possibilities, grouping certain particle trajectories into single field histories (that need not obey any particular field equations), a realistic interpretation of the Feynman path integral does appear to be possible.  Such an interpretation is not available to Hamiltonian-based formulations of quantum theory, not only because of the role of the future boundary constraint, but also because of the Lagrangian ``all at once'' style of analysis of entire histories.  The benefit of such a perspective, if fully developed, would be a reformulation of quantum theory in terms of generally covariant, spacetime-based quantities -- perhaps allowing the path integral to live up to its original promise, after all.

\section*{Acknowledgments}
The author would like to thank R. Sorkin and F. Dowker for the presentations and conversations that motivated this work (Cambridge, July 2014).  Further thanks go to the anonymous Referees, and also to editor Eliahu Cohen and Peter Evans, who all provided crucial feedback on earlier versions of this paper, leading to significant technical and explanatory improvements.

\end{document}